\documentclass[aps,11pt,final,notitlepage,oneside,onecolumn,nobibnotes,nofootinbib,superscriptaddress,noshowpacs,centertags]{revtex4}

\RequirePackage[cp1251]{inputenc}

\usepackage{amsmath,amssymb,epsfig,placeins,subfigure,wrapfig,indentfirst}

\input epsf
\begin{document}

\begin{center}
\Large {\bf Image of another universe being observed\\
through a wormhole throat}\\ Alexander Shatskiy
\footnote{Astrospace Center of the P.N. Lebedev Physics Institute
of Russian Academy of Sciences, 117997, ul.~Profsoyuznaya 84/32,
Moscow, 117997 Russia. E-mail: shatskiy@asc.rssi.ru.
Phone:~+7~(495)~333~33~66.}
\end{center}

\begin{center}\Large
Abstract
\end{center}
$\qquad$ We consider a technique of calculating deflection of the
light passing through wormholes (from one universe to another). We
find fundamental and characteristic features of electromagnetic
radiation passing through the wormholes. Making use of this, we
propose new methods of observing distinctive differences between
wormholes and other objects as well as methods of determining
characteristic parameters for different wormhole models.

\section{Introduction}
\label{s0}

Lately, relativistic astrophysics has seen an increasing interest
in papers where solutions with traversable wormholes (WHs) are
discussed. This interest is caused, among other things, by
projecting and constructing high-precision radiointerferometers
which will make it possible to discriminate WHs from black holes.

In this paper we consider a static and spherically symmetric WH
solution which only slightly differs from that for the extremal
charged Reissner-Nordstr\"{o}m black hole (see~\cite{3} or
\cite{7}). As is shown in \cite{2}, if the equation-of-state
parameters of matter change their values the WH solution can
smoothly turn into the extremal Reissner-Nordstr\"{o}m solution
with a horizon and the WH stops being traversable.

WH sustains its exotic properties thanks to phantom matter
violating the null energy condition\footnote{This condition means
satisfying the following inequality (NEC):
${T_{ik}\zeta^i\zeta^k>0}$, $T_{ik}$ being the energy-momentum
tensor of the matter and $\zeta^i$ -- the null 4-vector of photon.
Physically, the exoticness of the phantom matter violating the NEC
leads to the possibility of choosing a reference frame where
observed energy density is negative.} and surrounding a WH's
throat -- see, for example,~\cite{1} or \cite{2}.

What can be practically interesting is the WH that only slightly
differs from the extremally (whether electrically or magnetically)
charged Reissner-Nordstr\"{o}m black hole with the charge $q$. In
this particular case the amount of the phantom matter can be taken
arbitrarily small.

\section{The Einstein equations}
\label{s1}

We take the metric tensor for a static and spherically symmetric
WH in the following form:
\begin{equation}
ds^2= e^{2\phi (r)}\cdot [dt^2] - e^{\lambda (r)}\cdot [dr^2] -
r^2\cdot \left( [d\theta^2] + \sin^2\theta\, [d\varphi^2]\right)
\, . \label{metric}\end{equation} Let us consider the matter with
a linear equation-of-state where the relation between the energy
density ${\bf\varepsilon}$ and both the longitudinal (along
radius) and transverse (perpendicular to radius) pressures
($p_\parallel$ and $p_\perp$, respectively) is determined by a
constant factor ${1+\delta}$:
\begin{equation}
1+\delta =-p_\parallel /\varepsilon=p_\perp /\varepsilon \, .
\label{1-1}\end{equation} Using the notations
\begin{equation}
x=r/q \, ,\quad e^{-\lambda}\equiv 1-\frac{y(x)}{x}\, ,\quad \xi
(x)\equiv 8\pi\varepsilon q^2\, , \quad z(x)\equiv(1-1/x)\, .
\label{1-2}\end{equation} and introducing the prime for
derivatives with respect to $x$ allows one to write down the
Einstein equations\footnote{Their derivation is available, for
example, in \cite{4} (task 5 to \S 100).}:
\begin{eqnarray}
8\pi \varepsilon r^2=\xi x^2 = -e^{-\lambda}\left(1- x\lambda'
\right)+1\, ,
\label{En1}\\
8\pi p_\parallel r^2= -(1+\delta) \xi x^2 =
e^{-\lambda}\left(1+2x\phi' \right) -1 \, ,
\label{En2}\\
8\pi p_\perp r^2=(1+\delta) \xi x^2 = e^{-\lambda}\left( x^2\phi''
+x^2{\phi'}^2-x\lambda'/2 - x^2\phi'\lambda'/2+x\phi' \right) \, .
\label{En3}\end{eqnarray} The Einstein equations comprise the
Bianchi identities which in this case have the form:
\begin{eqnarray}
(1+\delta)(\ln\xi)' +  4(1+\delta)(\ln x)' + \delta\phi' = 0
\label{Bianki}\end{eqnarray} Denoting ${x_0\equiv y(x_0)}$ the
coordinate of the WH throat, we represent eqs. (\ref{En1}),
(\ref{En2}), and (\ref{Bianki}), in the following convenient form:
\begin{eqnarray}
y(x) = x_0+\int\limits_{x_0}^x \xi\, x^2\, dx
\label{En1-2}\\
\left(\ln\left[\xi x^4\right]\right)' = {\delta\over
2(1+\delta)}\cdot \frac{\xi x (1+\delta)-y/x^2}{1-y/x}
\label{En2-2}\\
\exp [\phi (x)] = \left[\xi x^4\right]^{-(1+\delta)/\delta}
\label{Bianki2}
\end{eqnarray}
If $\delta=0$, this solution turns into the extremal
Reissner-Nordstr\"{o}m solution which has a horizon:
\begin{eqnarray}
y_{_{[\delta=0]}}(x)=2-1/x\, ,\quad \xi_{_{[\delta=0]}}(x)=1/x^4\,
, \quad \exp[\phi_{_{[\delta=0]}}(x)]=1-1/x \, . \label{delta0}
\end{eqnarray}
Let us obtain the explicit analytical form of the solution to the
first order in the small correction $\delta$. In the linear
approximation with respect to $\delta$ eq. (\ref{En2-2}) takes the
form:
\begin{equation}
{\partial \ln (\xi\, x^4) \over \partial x} = -\delta \cdot
{\partial \ln (z) \over \partial x} \label{1-4}\end{equation}
Taking into account that, at infinity, the value of $\xi$ must be
equal to that of $\xi_{_{[\delta=0]}}$, we obtain the approximate
solution:
\begin{equation}
\xi\, x^4 = z^{-\delta} \, , \quad y(x) = x_0+
\frac{z^{1-\delta}-z_0^{1-\delta}}{1-\delta} \quad \left(z_0\equiv
1-\frac{1}{x_0}\right)\, ,\quad \exp (\phi) = z^{1+\delta} \, .
\label{approx1}\end{equation} The relation between the throat
radius $x_0$ and $\delta$ follows from the fact that the
asymptotics of the functions $e^{-\lambda}$ and $e^{2\phi}$ must
become equal as ${x\to\infty}$:
\begin{equation}
x_0+\frac{1-z_0^{1-\delta}}{1-\delta} \to 2(1+\delta)
\label{1-5}\end{equation} This is a transcendental equation
yielding the throat radius. The detailed analysis of the equation
gives the asymptotics:
\begin{equation}
\lim_{\delta\to 0}\delta = (x_0-1)^2, \label{1-6}\end{equation}
with the parameter $\delta$ being defined by the equation-of-state
of the matter in the WH.

Strictly speaking, the expansion into a series with respect to
$\delta$ is not quite well-posed. A straightforward substitution
of relations (\ref{approx1}) in eqs. (\ref{En2}) and (\ref{En3})
clearly demonstrates this. But the WH solution can be looked for
in an inverse manner, viz. first, one can use expressions
(\ref{approx1}) as the WH metrics components $e^{2\phi}$ and
$e^\lambda$ and then find expression for $p_\parallel$ and
$p_\perp$ from eqs. (\ref{En2}) and (\ref{En3}). This last
approach is no worse defined than just solving the Einstein
equations.

\section{Light passing through the throat}
\label{s2}

Let the other universe contain ${\bf N}$ stars with equal
luminosities and suppose ${\bf N>>>1}$. Let all the stars be
homogeneously distributed over the celestial sphere in the other
universe.

An observer in our Universe who is looking at the stars in the
other universe through the WH throat sees them inhomogeneously
distributed over the throat. This is because of the fact that the
WH throat refracts and distorts the light of these stars. The
distortion will obviously be spherically symmetric with the throat
center being the symmetry center.

Now let the observer look only at the fraction of the stars seen
in the thin ring with the center coinciding with the throat
center, the ring radius being ${\bf h}$ and its width -- ${\bf
dh}$. Hence, the observer surveys the solid angle ${\bf d\Omega}$
of the other universe and, moreover, ${\bf d\Omega = 2\pi
|\sin\theta |\, d\theta}$. Here, ${\bf \theta(h)}$ is the
deflection angle of light rays passing through the WH throat
measured relative to the rectilinear propagation\footnote{By
convention, the rectilinear propagation means the trajectory
passing through the center of the WH throat.}. Since the total
solid angle equals ${\bf 4\pi}$, the observer can see ${\bf dN=N\,
d\Omega /(4\pi)}$ stars in the ring\footnote{Since the light
deflection angle $\theta$ can exceed $\pi$, the total solid angle
turns out to be more than ${4\pi}$. This change, however, reduces
to another constant (instead of ${4\pi}$) and does not affect the
final result.}. Furthermore, the apparent density of the stars
(per unit area of the ring ${\bf dS=2\pi h\, dh}$) is ${\bf
J=dN/dS}$. We, therefore, obtain:
\begin{equation}
J(h)=\frac{N\,  |\sin\theta |}{4\pi h}\cdot \frac{d\theta}{dh}
\label{2-1}\end{equation}

In the paper \cite{2} the dependance $\theta (h)$ was obtained:
\begin{equation}
\theta (h)= 2\int\limits_0^{1/x_0}\, {\tilde h\over
\sqrt{(1-\eta\, y)(e^{-2\phi} - \eta^2\, \tilde h^2)}} \, d\eta \,
, \label{2-2}\end{equation} where the notations ${\eta\equiv 1/x}$
and ${\tilde h\equiv h/q}$ were used. This yields
\begin{equation}
\frac{d\theta}{dh} = \frac{2}{q}\int\limits_0^{1/x_0}\,
{e^{-2\phi} \over \sqrt{1-\eta\, y}\cdot\left[e^{-2\phi} -
\eta^2\, \tilde h^2\right]^{3/2} } \, d\eta \, .
\label{2-3}\end{equation}

Taking advantage of formulae (\ref{approx1}), (\ref{2-1}),
(\ref{2-2}), and (\ref{2-3}), one can find the expression for the
apparent density of the stars ${\bf J(h)}$ in the wormhole.

Note that formula (\ref{2-2}) also gives the maximum possible
impact parameter ${\bf h=h_{max}}$ which still allows the observer
to see the stars of the other universe. This parameter corresponds
to a zero of the second factor in the radicand in (\ref{2-2}).
Namely, ${\bf h_{max}}$ is equal to the least possible value of
the function ${\bf e^{-\phi}/\eta}$. Having conducted trivial
inquiry, we obtain
\begin{equation}
\tilde h_{max}=4\cdot 2^\delta \approx 4 \, .
\label{2-3}
\end{equation}
as ${\bf\delta\to 0}$.

\begin{figure*}
\includegraphics[width=0.99\textwidth]{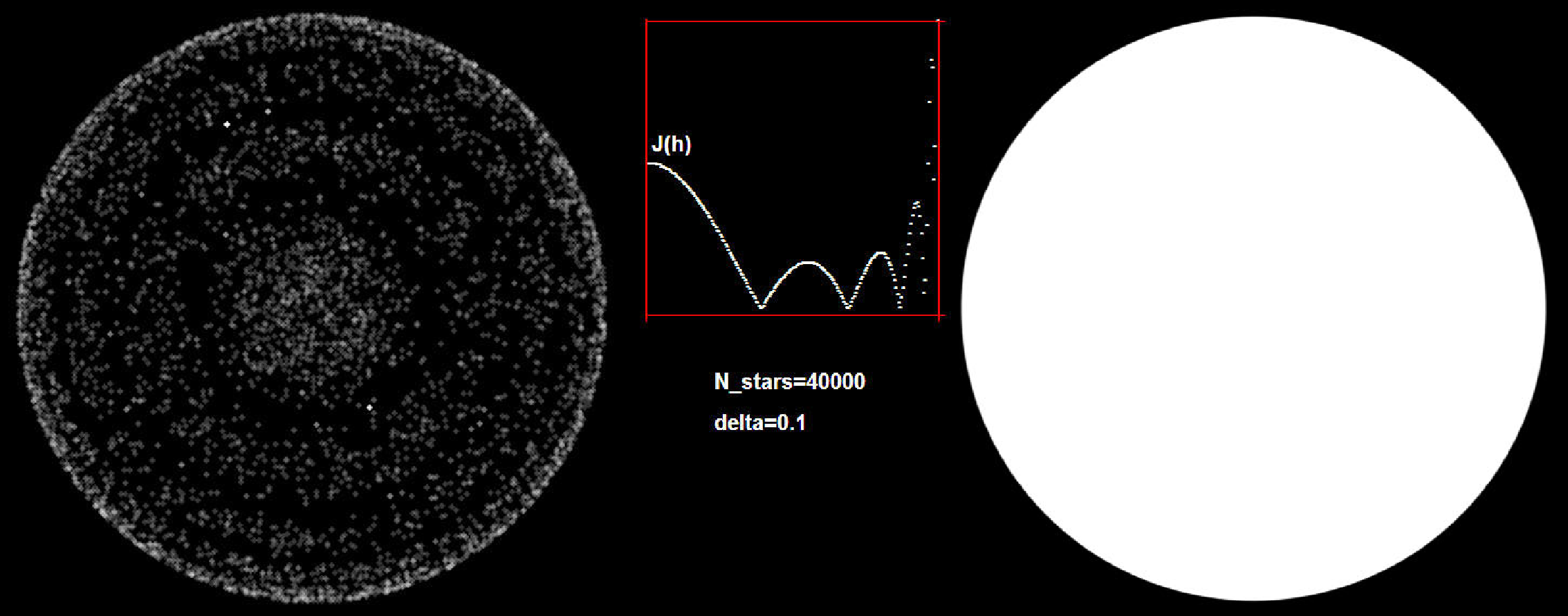} \caption{{
The left panel is the apparent image of the sky of the other
universe as being seen through the WH throat as ${N_{stars}=40\,
000}$. The middle panel shows the appropriate dependance ${J(h)}$
when ${h \in (0,h_{max})}$ and ${\delta =0.1}$. The right panel is
the apparent image of the sky of the other universe at
${N_{stars}\to\infty}$.}} \label{R1}
\end{figure*}

The distortion of the light rays that had passed through the WH
throat is caused not only by re-distributing of the star density,
but also by changes in their apparent brightness. Namely, as the
impact parameter ${\bf h}$ increases the stellar brightness
changes. This is because of the fact that as the radius ${\bf h}$
of the ring, through which the star light passes, increases, an
element of the solid angle where this light scatters changes as
well. The respective change in the stellar brightness is
proportional to the quantity ${\bf \kappa =dS/d\Omega}$.
Therefore, the total brightness of all the stars seen on unit area
of the above-mentioned ring is ${\bf dN\cdot \kappa /dS}$.

Thus, we obtain that as ${\bf N\to\infty}$ the apparent brightness
of the WH's part inside its throat does not depend on impact
parameter and, regardless of which WH model we use, the WH looks
like a homogeneous spot in every wavelength range.

We point out in this regard that the conclusion in the paper
\cite{Sh3} about inhomogeneous observation of the light from the
other universe is false. In that article correct mathematical
expressions were misinterpreted.

\section{Conclusion}
\label{s-zakl}

In spite the result obtained stating that the light distribution
in the WH throat is homogeneous for each WH model, it is worth
noting that in the real universe the number of visible stars is
finite, though big. This implies that if angular resolution of the
observer's instrument in our Universe is high enough they will be
able to discover the changing star density in the throat ${\bf
J(h)}$. The left panel of Fig.~\ref{R1} shows this plot for
${\delta =0.001}$. Sharp minima on the plot correspond to zeros of
the sine in expression (\ref{2-1}). This is because at
sufficiently large impact parameters the light rays are deflected
by large angles (${\theta>\pi}$) so that in the vicinities of the
points ${\theta =\pi n}$ abrupt declines in distribution arise.
But near these declines the observed stellar brightness tends to
infinity (lensing), which ultimately provides the (on average)
uniform light flow over the WH throat (see the right panel of
Fig.~\ref{R1}).

Positions of the declines depend on the value of $\delta$. Hence,
registering them makes it possible to determine the
equation-of-state parameters of the WH matter and features of the
WH model (which is highly analogous to processing the light
spectra).

Thus, in this paper we have proposed a technique of calculating
the deflection of light passing through wormholes as well as
methods of observing distinctive features of specific WH models.

\section{Acknowledgements}
\label{s-end}

Finally, I would like to express my gratitude to Vladimir Strokov,
N.S.~Kardashev and I.D.~Novikov for fruitful discussion of this
paper and their valuable remarks.

\bigskip
This work is supported by the Russian Foundation for Basic
Research (project codes: 07-02-01128-a, 8-02-00090-a), scientific
schools: NSh-626.2008.2, Sh-2469.2008.2, and by the program Origin
and Evolution of Stars and Galaxies 2008 of Russian Academy of
Sciences.


\end{document}